\documentstyle[aps]{revtex}
\begin{document}

\begin{center}
{\Large Spin-Peierls Dimerization of a $s=\frac 12$ Heisenberg
Antiferromagnet on Square Lattice}\\\vspace{1cm}

Aiman Al-Omari and A. H. Nayyar$^{*}$\\

Department of Physics, \\Quaid-i-Azam University, \\Islamabad, Pakistan
45320.\\

\vspace{1cm}(June 25, 1997)\vspace{2.5cm}

{\bf Abstract}\vspace{0.7cm}
\end{center}

Dimerization of a spin-half Heisenberg antiferromagnet on a square lattice
is investigated by taking unexpanded exchange couplings. Several dimerized
configurations are considered some of which are shown to have lower ground
state energies than others. In particular, the lattice deformations
resulting in alternate strong and weak couplings along both the principal
axes of a square lattice are shown to result in a larger gain in energy. In
addition, a `columnar' configuration is shown to have a lower ground state
energy and a faster increase in the energy gap parameter than a `staggered'
configuration. The inclusion of unexpanded exchange coupling leads to a
power law behaviour for the magnetic energy gain and the spin-Peierls gap,
which is qualitatively different from that reported earlier. Instead of
varying as $\delta ^x$, the two quantities are shown proportional to $\delta
^\nu /\left| \ln \delta \right| .$ It is thus proposed that the logarithmic
correction, that was regarded as an outcome of including umklapp processes
for small distortions, can also be a direct consequence of using an
un-truncated spin-spin exchange interaction. The calculations, which employ
the coupled cluster method, lead to a conclusion result that the
dimerization of a spin-half Heisenberg antiferromagnet on a sqaure lattice
is unconditional. \vspace{1cm}

\newpage

\section*{I. Introduction}

It is known that dimerization lowers the ground state energy of a spin-half
isotropic Heisenberg antiferromagnet\cite
{soos,jacobs,cross,fields,matsuyama,bonner,soos1,spronken,guo,chitra}. In
other words, the system stands to gain energy by such lattice deformations
that render it dimerized with alternate weaker and stronger bonds between up
and down spins on neighboring sites. On the other hand the lattice
distortions cost energy and it is the net energy balance that would
determine whether the gain in magnetic energy is large enough to affect the
spin-Peierls transition through dimerization. In a phenomenological theory,
this is usually seen in terms of an exponent showing the dependence of
magnetic and elastic energies on a parameter $\delta $, describing the
extent of lattice deformations. It is related to the displacement of the $i$%
th atom through $u_i=\frac 12(-1)^i\delta $. The spin-dimer formation is
usually described by the Hamiltonian 
\begin{equation}
H=J\sum_i[1+(-1)^i\delta ]{\bf S}_i\cdot {\bf S}_{i+1}
\end{equation}
with alternate stronger and weaker bonds $J(1+\delta )$ and $J(1-\delta )$.

Since $0\leq \delta \leq 1$, and since elastic energies go typically as $%
\delta ^2$, therefore if the magnetic energy gain varies with $\delta $ with
an exponent less than 2 then in the limit $\delta \rightarrow 0$, the gain
would overwhelm the cost and a spontaneous and {\it unconditional}
dimerization should occur. If, however, the exponent is equal to or greater
than 2 then dimerization may occur, {\it conditional} to the details of the
interaction parameters.\\

Such aspects as these have been studied extensively in Heisenberg
antiferromagnetic chains in which the critical exponent has been shown to
favor an unconditional spin-Peierls transition, as summarized in Table 1.
This aspect has also been revealed by experiments on quasi-one dimensional
Heisenberg antiferromagnet CuGeO\cite{hase,nishi,fujita}.\\

Table 1 lists the scaling laws proposed for the magnetic energy gain in one-
dimensional antiferromagnets both in the small $\delta $ (near critical)
regime as well as the far critical regime ($0\leq \delta \leq 1$). Of
particular interest to us is the logarithmic correction for small $\delta $,
which is reported to be a result of including umklapp processes\cite
{spronken}. We find, however, that if, instead of the approximated form of
the spin-spin exchange coupling, $J(1\pm \delta )$, its full form 
\begin{equation}
J(r_{ij})=\frac J{\left| r_{ij}\right| }  \label{fullj}
\end{equation}
is used then the magnetic energy gain comes to vary with $\delta $ as $\frac{%
\delta ^\nu }{\left| \ln \delta \right| }$, exactly as calculated by the
various methods shown in Table 1. Additionally, it is easy to see that this
scaling behaviour will not be confined to the near critical region, but will
operate over the entire range of $\delta $. The full form of the interaction
includes the often-used truncated forms in an obvious way: when the distance
between a pair of spins decreases from $a$ to $a(1-\delta )$, the exchange
coupling is taken to increase from $J$ to approximately $J(1+\delta )$, etc.
In what follows, we shall use for exchange interaction the form in Eq.(\ref
{fullj}).\\

The situation in two-dimensions is a little more involved because of the
additional factor of frustration due to a competing antiferromagnetic second
neighbour interaction which can in principle destroy any LRO of the Neel
type as well as the possibility of dimerization. Much of the study of
two-dimensional Heisenberg antiferromagnet has therefore remained focused on
the destruction of order by frustration.\\

The matter of frustration aside, a simple dimerization of a square lattice
is interesting in its own right because the lattice distortions can take
place in more than one way, each one of the possible configurations giving a
different dependence of the ground state energy on the dimerization
parameter. Studies of dimerization of a spin-half Heisenberg antiferromagnet
on a square lattice have mostly considered a columnar configuration,
illustrated in Fig.1(a)\cite{rokhsar,tang,read,xu,katoh,feiguin}.{\small \ }%
That is to say, the lattice deformations and the consequent spin-singlet
pair formations are taken to occur only along one of the square axes, say
the $x$-axis. The nearest neighbour distance along that axis is taken to
vary alternately as $a(1-\delta )$ and $a(1+\delta )$, while that between
neighbours in the perpendicular ($y$-) direction remains $a$. It has been
shown that, like the chains, this configuration also gives rise to a
continuous reduction of ground state energy with $\delta $.\\

One can think of other possible configurations of two-dimensional
dimerization which are different from this one in the sense that they allow
for lattice distortions or changed spin-spin couplings along the $y$%
-direction also. We shall propose a few such configurations in the
following, three of which are shown in Figs.1(b)-(d).\\

In Fig.(b), the lattice deformations along the $x$-direction is alternated
as in Fig.(a), but the sequence of alternations is itself alternated as one
goes along the $y$-direction. To distinguish it from the columnar
dimerization, we call it staggered dimerization\cite{leung}. It makes the
exchange coupling along the $y$-direction also dependent upon the
dimerization parameter $\delta $. While the coupling along the $x$-direction
is alternately $\frac J{1-\delta }$ and $\frac J{1+\delta }$, it is
uniformly $\frac J{\sqrt{(1+\delta ^2)}}$ along the $y$-direction.\\

In contrast to the configurations (a) and (b), those in Figs.1(c) and (d)
allow for simultaneous dimerization along both $x$- and $y$-directions in
the plane. The difference between (c) and (d) is the same as that between
(a) and (b): configuration (c) is columnar and (d) is staggered. The former
is called plaquette configuration\cite{tang,leung}. These four
configurations of a dimerized square lattice consisting of $N$ spins are
therefore characterized by the following nearest neighbour interactions.\\

{\bf Configuration (a)}\\

$J_{x,\lambda }=\frac J{(1+\lambda \delta )}\simeq J(1-\lambda \delta ),$ $%
\lambda =\pm 1$ \\

$J_y=J.$\\

That is to say, the dimerized configuration is described by the Hamiltonian

\begin{eqnarray}
H=J\sum_{i,j}^{\sqrt{N}}\left[ \frac 1{\left( 1+(-1)^i\delta \right) }{\bf S}%
_{i,j}\cdot {\bf S}_{i+1,j}+{\bf S}_{i,j}\cdot {\bf S}_{i,j+1}\right]
\label{h1}
\end{eqnarray}

{\bf Configuration (b)}\\

$J_{x,\lambda }=\frac J{(1+\lambda \delta )}\simeq J(1-\lambda \delta
),\lambda =\pm 1$ \\

$J_y=\frac J{\sqrt{1+\delta ^2}}\simeq J(1-\frac{\delta ^2}2)$\\

and the Hamiltonian is given by

\begin{eqnarray}
H=J\sum_{i,j}^{\sqrt{N}}\left[ \frac 1{\left( 1+(-1)^{i+j}\delta \right) }%
{\bf S}_{i,j}\cdot {\bf S}_{i+1,j}+\frac 1{\sqrt{1+\delta ^2}}\,\,{\bf S}%
_{i,j}\cdot {\bf S}_{i,j+1}\right]  \label{h2}
\end{eqnarray}

{\bf Configuration (c)}\\

$J_{x,\lambda }=\frac J{(1+\lambda \delta )}\simeq J(1-\lambda \delta ),$ $%
\lambda =\pm 1$ \\

$J_{y,\lambda }=\frac J{\left( 1+\lambda \delta \right) }\simeq J\left(
1+\lambda \delta \right) $\\

with the Hamiltonian

\begin{eqnarray}
H=J\sum_{i,j}^{\sqrt{N}}\left[ \frac 1{\left( 1+(-1)^i\delta \right) }{\bf S}%
_{i,j}\cdot {\bf S}_{i+1,j}+\frac 1{\left( 1+(-1)^j\delta \right) }{\bf S}%
_{i,j}\cdot {\bf S}_{i,j+1}\right]  \label{h3}
\end{eqnarray}

{\bf Configuration (d)}\\

$J_{x,\lambda }=\frac J{(1+\lambda \delta )}\simeq J(1-\lambda \delta ),$ $%
\lambda =\pm 1$\\

$J_{y,\lambda }=\frac J{\sqrt{\left( 1+\lambda \delta \right) ^2+\delta ^2}}%
\simeq J\left( 1-\lambda \delta -(1-\frac{\lambda ^2}2)\delta ^2\right) $\\

and the Hamiltonian

\begin{equation}
H=J\sum_{i,j}^{\sqrt{N}}\left[ \frac 1{\left( 1+(-1)^{i+j}\delta \right) }%
{\bf S}_{i,j}\cdot {\bf S}_{i+1,j}+\frac 1{\sqrt{\delta ^2+\left(
1+(-1)^j\delta \right) ^2}}\,\,{\bf S}_{i,j}\cdot {\bf S}_{i,j+1}\right]
\label{h4}
\end{equation}

We would like to investigate the four configurations in order to see (i)
which one of these leads to the largest gain in magnetic energy as the
dimerization sets in, (ii) whether such a dimerization is conditional or
otherwise, and (iii) if the use of un-truncated exchange coupling leads to a
single scaling law valid for the entire range of $\delta $.\\

Being a prototype of the superconducting phase transition\cite{pytte}, the
appropriate order parameter for spin-Peierls transition must also be the
energy gap between the ground and excited states - the energy required to
break a singlet pair. We shall therefore look at the critical behaviour of
the system under dimerization in terms of the $\delta $-dependence of the
energy gap parameter.\\

A number of methods can be chosen for this purpose. Spin wave theory, either
modified through Takahashi constraint of zero magnetization or a
Hartree-Fock approximated non-linear theory, is known to give surprisingly
good results for spin-half Heisenberg antiferromagnet. Or, a spin wave
theory in the spinless fermionic representation through Jordan-Wigner
transformations takes care of fermionic correlations among the $s=\frac 12$
spins. Coupled cluster method has also been extensively, and successfully,
used for spin-half Heisenberg antiferromagnet in one and two space
dimensions.\\

The first two methods belong to the class of mean field theories and hence
are not expected to be very reliable when it comes to determining critical
exponents. The coupled cluster method, on the other hand, is a perturbation
method in which increasingly higher order correlations can, in principle, be
incorporated at will, and which has been shown to give satisfactory results
even in the lower orders of perturbation. We believe that the coupled
cluster method must be sufficiently good to see if the energetics allow a
spin-Peierls transition to the alternative configurations proposed here.\\

\section*{II. Application of the coupled-cluster method}

In the coupled cluster method it is first necessary to define a ket state
starting from a model state $\mid \phi >,$ which in our case is the Neel
state. The exact ground state $\mid \Psi >$ of the system can then be
postulated as 
\begin{equation}
\mid \Psi >=e^{{\cal S}}\mid \phi >
\end{equation}
where ${\cal S}$ is the correlation operator defined for an $N$ particle
system as 
\begin{eqnarray}
{\cal S} &=&\sum_n\,{\cal S}_n\,\, \\
\mbox{with     }{\cal S}_n &=&\sum_{i_1...i_n}{\sf S}%
_{i_1,....,i_n}C_{i_1}^{\dagger }C_{i_2}^{\dagger }\cdot \cdot \cdot \cdot
\cdot \cdot C_{i_n}^{\dagger }
\end{eqnarray}
and $C_i^{\dagger }$ is the creation operator defined with respect to the
model state. The ground state energy can then be found as the eigenvalue of
the Hamiltonian in the proposed ground state\\

$He^{{\cal S}}\mid \phi >=E_ge^{{\cal S}}\mid \phi >$.\\

Taking inner product with $<\phi \mid e^{-{\cal S}}$ gives\\

$E_g=<\phi \mid e^{-{\cal S}}He^{{\cal S}}\mid \phi >.$\\

The product $e^{-{\cal S}}He^{{\cal S}}$ can be written as a series of
nested commutators in the well-known expansion 
\begin{equation}
e^{-{\cal S}}He^{{\cal S}}=H+[H,{\cal S}]+\frac 1{2!}\left[ [H,{\cal S}],%
{\cal S}\right] +\cdot \cdot \cdot \cdot \cdot \cdot \cdot \cdot \cdot
\end{equation}
where in the present case the series terminates after the fourth term.\\It
is usually easier to deal with the $s=\frac 12$ Heisenberg Hamiltonian by
applying a rotation of 180$^{\circ }$ to the up spin sublattice such that in
it $S_x\rightarrow -S_x,$ $S_y\rightarrow S_y$ and $S_z\rightarrow -S_z$.
The fermionic character of the spin operators is then preserved by
expressing them in terms of Pauli matrices: $S^j=\frac 12\sigma ^j,$ $%
j=x,y,z $\cite{bishop,bishop1}. A general expression for the nearest
neighbour spin Hamiltonian in 2D is then 
\begin{equation}
H=-\frac J4\sum_{{\bf l,\rho }}\left( 2(\sigma _{{\bf l}}^{+}\sigma _{{\bf %
l+\rho }}^{+}+\sigma _{{\bf l}}^{-}\sigma _{{\bf l+\rho }}^{-})+\sigma _{%
{\bf l}}^z\sigma _{{\bf l+\rho }}^z\right) ,
\end{equation}
where ${\bf \rho }$ is a vector to the four nearest neighbours.
Correspondingly, the string operator ${\cal S}_n$ can now be defined as 
\begin{equation}
{\cal S}_{2n}=\frac 1{(n!)^2}\sum_{{\bf i}_1{\bf ....i}_n}\sum_{{\bf j}_1%
{\bf ....j}_n}{\sf S}_{{\bf i}_1{\bf ...i}_n;\,\,{\bf \,j}_1{\bf ....j}%
_n}\sigma _{{\bf i}_1}^{+}\sigma _{{\bf i}_2}^{+}\cdot \cdot \cdot \sigma _{%
{\bf i}_n}^{+}\sigma _{{\bf j}_1}^{+}\sigma _{{\bf j}_2}^{+}\cdot \cdot
\cdot \sigma _{{\bf j}_n}^{+},
\end{equation}

where subscripts ${\bf i}$ and ${\bf j}$ distinguish between sites on the
two sublattices. We note that for spin half $(\sigma _{{\bf l}%
}^{+})^2=(\sigma _{{\bf l}}^{-})^2=0.$ Truncation of the summation up to the
desired level gives rise to different schemes of approximation. Taking
interaction only between the spins on adjacent sites gives the so-called SUB$%
_{2-2}$ scheme. Including interactions with the second and fourth
neighboring sites gives what is termed as SUB$_{2-4}$ scheme. And taking the
previous two schemes including interaction among the four adjacent sites
give us what has been termed as local SUB$_4$, or LSUB$_4$ for short. Each
one of these approximations accounts for a different order of perturbation
calculation, and takes into account a different order of inter-particle
correlations. It has been noted that LSUB$_4$ is a sufficiently good
approximation for calculating the ground state properties of a spin-half
Heisenberg system \cite{bishop1}.\\

Consider a general case: a Hamiltonian which has four different coupling
constants for nearest neighbour interactions in two space dimensions. It can
be written as

\begin{eqnarray}
H=-\frac 14\sum_{i,j}^{\sqrt{N}/2}\sum_{\lambda =\pm 1}\left[ J_{x,\lambda
}\,\,\,{\bf \sigma }_{2i,j}\cdot {\bf \sigma }_{2i+\lambda ,j}+J_{y,\lambda
}\,\,\,{\bf \sigma }_{i,2j}\cdot {\bf \sigma }_{i,2,j+\lambda }\right] .
\end{eqnarray}

Here $i$ and $j$ are the two components of the site indices on a square
lattice. The correlation operators in the LSUB$_4$ scheme are defined as\\
\begin{eqnarray}
{\cal S}_2 &=&\sum_i\left[ a_1\,\,\,\sigma _{2i,j}^{+}\sigma
_{2i+1,j}^{+}+b_1\sigma _{2i,j}^{+}\sigma _{2i-1,j}^{+}\,\,\,+c_1\sigma
_{i,j}^{+}\sigma _{i,j+1}^{+}+d_1\sigma _{i,j}^{+}\sigma _{i,j-1}^{+}\right]
\nonumber \\
{\cal S}_3 &=&\sum_i\left[ a_3\sigma _{2i,j}^{+}\sigma
_{2i+3,j}^{+}\,\,\,+b_3\sigma _{2i,,j}^{+}\sigma
_{2i-3,j}^{+}\,\,\,+c_3\sigma _{i,j}^{+}\sigma _{i,j+3}^{+}+d_3\sigma
_{i,j}^{+}\sigma _{i,j-3}^{+}\right]  \nonumber \\
{\cal S}_4 &=&\sum_i\left[ f\,\,\prod_{\nu =0}^3\sigma _{2i+\nu
,j}^{+}+g\,\prod_{\nu =0}^3\sigma _{2i-\nu ,j}^{+}+h\,\prod_{\nu =0}^3\sigma
_{i,j+\nu }^{+}\,+l\,\,\prod_{\nu =0}^3\sigma _{i,j-\nu }^{+}\right]
\end{eqnarray}

In these equations, the coefficients $a_{1,}b_{1,}$etc., are various forms
of the coefficient {\sf S}$_{{\bf i}_1{\bf ...i}_n\,;\,\,{\bf j}_1{\bf ....j}%
_n}$ in the expressions for ${\cal S}_{2n}$. The ground state energy within
the LSUB$_4$ approximation comes out to be

\begin{equation}
E_g=-\frac 14\left[ J_{x,+1}\,\,\,\left( \frac 14+a_1\right) +J_{x,-1}\,\,\,(%
\frac 14+b_1)+J_{y,+1}\,\,\,(\frac 14+c_1)+J_{y,-1}\,\,\,(\frac 14%
+d_1)\right]
\end{equation}

The coefficients $a_1,a_2,\cdot \cdot \cdot ,l$ are obtained as solutions of
a set of coupled nonlinear equations. These equations arise from the fact
that such matrix elements as $<\phi \mid {\bf O}$ $e^{-{\cal S}}He^{{\cal S}%
}\mid \phi >$ are all zero when the operator ${\bf O}$ is any product of
creation operators, particularly if it is one of the operator products in
the correlation operator ${\cal S}$ above.

\begin{eqnarray}
\ &<&\sigma _{i,2j}^{-}\sigma _{i,2j+\nu }^{-}e^{-{\cal S}}He^{{\cal S}}>=0;%
\hspace{3.5cm}\nu =\pm 1,\pm 3  \nonumber \\
\ &<&\sigma _{2i,j}^{-}\sigma _{2i+\nu ,j}^{-}e^{-{\cal S}}He^{{\cal S}}>=0;%
\hspace{3.5cm}\nu =\pm 1,\pm 3  \nonumber \\
\ &<&\sigma _{2i,j}^{-}\sigma _{2i+\nu ,j}^{-}\sigma _{2i+2\nu ,j}^{-}\sigma
_{2i+3\nu ,j}^{-}e^{-{\cal S}}He^{{\cal S}}>=0;\hspace{1cm}\nu =\pm 1 
\nonumber \\
\ &<&\sigma _{i,2j}^{-}\sigma _{i,2j+\nu }^{-}\sigma _{i,2j+2\nu }^{-}\sigma
_{i,j+3\nu }^{-}e^{-{\cal S}}He^{{\cal S}}>=0;\hspace{1cm}\nu =\pm 1
\end{eqnarray}
where ${\cal S}={\cal S}_2+{\cal S}_3+{\cal S}_4$. These equations translate
into the following twelve equations for the unknown parameters:\\

$(a_1^2-1-2a_3b_1-2f)J_{x,+1}+(2a_1+2a_1b_1-2a_3b_3)J_{x,-1}{}=0$\\

$(2b_1+2a_1b_1-2a_3b_3)J_{x,+1}+(b_1^2-1-2a_1b_3-2g)J_{x,-1}=0$\\

${(2}a_3+2a_1a_3{)}J_{x,+1}{+(2}a_3-a_1^2+2a_3b_1-f{)}J_{x,-1}{=0}$ $\\$

$(2b_3-b_1^2+2a_1b_3-g)J{_{x,+1}+}(2b_3+2b_1b_3{)}J{_{x,-1}=0}$ $\\$

$(-2a_3b_1+2a_1f+a_3b_1^2-a_1a_3b_1{)}J{_{x,+1}}{+(}%
f-a_1^2+2b_1f+a_3g+2a_1a_3b_3{)}J{_{x,-1}=0}\\$

$%
(g-b_1^2+2a_1g+b_3f+2a_3b_1b_3)J_{x,+1}+(-2a_1b_3+2b_1g+a_1^2b_3-a_1b_1b_3)J_{x,-1}=0 
$ \\

$(c_1^2-1-2c_3d_1-2h)J_{y,+1}+(2c_1+2c_1d_1-2c_3d_3)J_{y,-1}=0$ \\

$(2d_1+2c_1d_1-2c_3d_3)J_{y,+1}+(d_1^2-1-2c_1d_3-2l)J_{y,-1}=0$\\

$(2c_3+2c_1c_3)J_{y,+1}+(2c_3-c_1^2+2c_3d_1-h)J_{y,-1}=0 $ \\

$(2d_3-d_1^2+2c_1d_3-l)J_{y,+1}+(2d_3+2d_1d_3)J_{y,-1}=0 $ \\

$(-2c_3d_1+2c_1h+c_3d_1^2-c_1c_3d_1)J_{y,+1}+
(h-c_1^2+2d_1h+c_3l+2c_1c_3d_3)J_{y,-1}=0 $ \\

$(l-d_1^2+2c_1l+d_3h+2c_3d_1d_3)J_{y,+1}
+(-2c_1d_3+2d_1l+c_1^2d_3-c_1d_1d_3)J_{y,-1}=0$\\

Setting all the coupling constants $J_\mu $ equal reduces the number of
equations from twelve to three and yields exactly the same equations as
obtained by others\cite{bishop,bishop1}. The two sets of six equations each
independently determines the six coefficients contained in each of them. As
expected, the equations are symmetric in some coefficients. The twelve
coefficients are to be evaluated by solving the above coupled equations
numerically for each of the configurations separately by substituting
appropriate values of $J_{x,\lambda }$ and $J_{y,\lambda }$.\\

To calculate the energy gap we shall construct the excited ket state $|\Psi
_e>$ in term of a linear excitation operator ${\bf X},$ which, operating on
the ground state $|\Psi _0>,$ takes the system to an excited state: $|\Psi
_e>={\bf X}$ $|\Psi _0>={\bf X}e^{{\cal S}}$ $|\phi >.$ This operator is
constructed as a linear combination of products of creation operators\cite
{bishop1} 
\begin{equation}
{\bf X}=\sum_nX_n
\end{equation}
with 
\begin{equation}
X_n=\sum_{{\bf j}_1{\bf ....j}_n}{\bf \chi }_{\,\,{\bf \,j}_1{\bf ....j}%
_n}\sigma _{{\bf j}_1}^{+}\sigma _{{\bf j}_2}^{+}\cdot \cdot \sigma _{{\bf j}%
_n}^{+}.
\end{equation}
The first excited state is obtained by the operator 
\begin{equation}
X_1=\sum_{{\bf j}}{\bf \chi }_{\,\,\,{\bf j}}\sigma _{{\bf j}}^{+}
\end{equation}
where ${\bf j}$ can be any site of the two sublattices. It is easily seen
that the first excitation energy is 
\begin{equation}
E_e=\frac 14\left( \frac 14+a_1+b_1+a_3+b_3\right) \left(
J_{x,+1}+J_{x,-1}\right) .
\end{equation}
The energy gap for a given $\delta $ is $\Delta (\delta )=E_e(\delta
)-\left| E_g(\delta )\right| .$ The order parameter for the spin-Peierls
transition is therefore $D(\delta )=\Delta (\delta )-\Delta (0).$ This is
the energy required to break a dimerized singlet pair.\\

\section{III. Results}

The ground state energy can now be calculated as a function of the
dimerization parameter $\delta $ . Previous calculations have invariably
taken spin-spin exchange couplings alternately as $J(1\pm \delta )$, which,
as mentioned above, is an expansion of the interaction in Eq.(\ref{fullj})
to order $\delta $, implying that the results are valid only in the limit $%
\delta \rightarrow 0$. We notice in our calculations that if all the
expansions are terminated beyond the order $\delta $ then the distinction
between configurations (a) and (b) [Eqs.(\ref{h1}, \ref{h2})] disappears. On
the other hand, if the expansion is taken to one order higher, then there
remains no way to distinguish between configurations (c) and (d), Eqs.(\ref
{h3},\ref{h4}). We must therefore either go to orders beyond $\delta ^2$ in
the expansion, or retain the interactions in their unexpanded form Eq.(\ref
{fullj}). We do the latter. An added advantage is that the results will then
be valid in the limit $\delta \rightarrow 1$.\\

Our calculations confirm that, like the chain, the ground state energy of
all the four configurations of a dimerized spin-half Heisenberg
antiferromagnet on a square lattice decreases with $\delta $. This is shown
in Figures 2, where $\varepsilon (\delta )-\varepsilon (0)$ is plotted
against $\delta $ for the proposed configurations. The conclusion is not new
for the simple columnar dimerization of Fig.1(a) and (c) \cite
{rokhsar,tang,read,xu,katoh,feiguin}. However, what is significant is that
the ground state energy goes down with $\delta $ more rapidly for
configurations (c) and (d), Figs. 1(c)-(d). In fact, Fig. 2 shows that the $%
\delta $-dependence is markedly different for the two types of dimerized
configurations: one in which dimerization takes place only along one axis,
and the other, in which it occurs along both the directions. The rate of
decrease is significantly higher for the latter. Also, for the entire range
of $\delta $ the columnar configurations lead to a greater gain in magnetic
energy than the staggered dimerization. The two observations put together
show that the plaquette configuration of Fig.1(c) is energetically the most
favourable state, as noted earlier\cite{tang}. Particularly in the complete
range of $\delta $ ($0\leq \delta \leq 1$), the plaquette configuration
stands out as the most preferred one, while there is hardly a discernible
difference among the other three.\\

It is worth pointing out here that the much simpler mean field methods of
spin wave theory - either in the bosonic representation through
Holstein-Primakoff transformations, or in the fermionic representation
through Jordan-Wigner transformations - yield very similar results. This has
been checked by us separately.\\

To see if the spin-Peierls transition setting in the Heisenberg
antiferromagnet on a square lattice is conditional or otherwise, we look at
the balance between the gain in magnetic energy through dimerization and the
cost in energy for elastic deformations; the latter going as $\delta ^2$.
Earlier calculations on a square lattice had obtained the exponent as 1\cite
{katoh} for columnar configuration (a), and as 2 for columnar as well as
staggered configurations\cite{tang,feiguin}. Our results must be different
from these because instead of $J(1\pm \delta )$, we take the unapproximated
exchange coupling $J(r_{ij})=\frac J{\left| r_{ij}\right| }$. However, they
must agree with the earlier results in the limit of small $\delta $.\\

We find that the gain in magnetic energy does not quite scale with $\delta $
as a simple power law: it scales as $\frac{\delta ^v}{\left| \ln \delta
\right| }$for all the four configurations. The exponent $\nu =1.5$ in the
range $0\leq \delta \leq 0.1$, and is equal to $1$ in the complete range $%
0\leq \delta \leq 1$. This is true for all the four configurations. That
Tang and Hirsch had obtained a simple power law with exponent 2 is easy to
understand if one notes that the difference between $\delta ^2$ and $\frac{%
\delta ^{1.5}}{\left| \ln \delta \right| }$ is difficult to discern in the
small $\delta $ regime, particularly while fitting to a curve.\\ 

Since $\frac{\delta ^v}{\left| \ln \delta \right| }$ for $\nu =1$ or $1.5$
is always larger than $\delta ^2$ in the entire range $0\leq \delta \leq 1,$
the spin-Peierls transition in a spin-half Heisenberg antiferromagnet on a
square lattice must be unconditional.\\

As shown in Table 1, the dimerization of an antiferromagnetic chain scales
as $\frac{\delta ^v}{\left| \ln \delta \right| }$, but only in the small $%
\delta $ regime (the near critical regime). In chains, the factor of $\frac 1%
{\left| \ln \delta \right| }$ is taken to serve as a correction due to
umklapp processes. This is not the case in a square lattice. In fact our CCM
results show that this may not be the case even in chains when the exchange
couplings in the dimerized state are taken as $\frac J{1\pm \delta }$
instead of the approximated $J(1\pm \delta )$. We find for chains that the
magnetic energy gain fits to $\frac{\delta ^v}{\left| \ln \delta \right| }$
in the entire range of $\delta $ rather than only in the range of small $%
\delta $. With the full exchange couplings, the coefficients come out for
the chain to be $\nu =\frac 23$ for $0\leq \delta \leq 1$, and $\nu =1.3$ $%
-1.6$ for $0\leq \delta \leq 0.1$. These give decent comparisons with the
numbers in Table 1.\\ 

The $\delta $ dependence of the order parameter for the four configurations
is shown in Fig.3. An obvious conclusion from these results is that the
dimerized state becomes more stabilized with increasing $\delta $. Like the
magnetic energy gain, the order parameter $D$ also scales with $\delta $ as $%
\frac{\delta ^v}{\left| \ln \delta \right| }$ with the same values for $\nu $
in both the small as well as full region of $\delta $. \\

The difference between the dimerization of a square lattice along only one
direction (Fig. 1(a) and (b)) and along both the directions (Fig. 1(c) and
(d)) is again markedly brought out in Fig. 3. Also the columnar
configurations again appear to be the preferred modes of dimerization over
the staggered configurations for having higher values of the order parameter
in the region of small $\delta $.\\

To summarize, we have studied the spin-Peierls dimerization of a spin-half
Heisenberg antiferromagnet on a square lattice taking unapproximated
exchange couplings starting from the ansatz $J(r_{ij})=\frac J{\left|
r_{ij}\right| }$. We have included different possibilities of dimerization.
The ground state energy decreases continuously with increasing dimerization
for all the proposed configurations, . Of the four configurations, those
with dimerization taking place simultaneously along both the principal
square axes have markedly lower ground state energies than those with
dimerization along only one of the axes. Also, those with columnar
dimerization have consistently lower energies than those with the staggered
dimerization. The spin-Peierls gap also corroborates the above conclusions.
It has also been shown that the magnetic energy gain as well as the
spin-Peierls gap scale with the dimerization parameter $\delta $ as $\frac{%
\delta ^v}{\left| \ln \delta \right| }$. It has also been asserted that this
form of scaling does not need to be a consequence of corrections due to
umklapp processes, and that the same can be said of the  spin-Peierls
transition in chains. As a result of this, it is concluded that  the
spin-Peierls transition in a spin-half antiferromagnet on a square lattice
is unconditional.\\ 

\noindent $*$ E-mail: aiman@physics.sdnpk.undp.org\\%
nayyar@physics.sdnpk.undp.org\\

\ 

\section*{Figure captions}

Figure 1: Four configurations for the dimerization of a square lattice. (a)
Columnar: the nearest neighbour coupling along the horizontal direction
alternates between $J(1-\delta )$ and $J(1+\delta )$, while that along the
vertical direction remains $J$. (b) staggered: like (a), the dimerization
occurs along one direction only, but the sequence of alternate couplings
itself alternates along the other direction. (c) Dimerization along both the
directions, making a plaquette of four nearest neighbour spins. (d) Again
dimerization along both the directions, but taken staggered along the
vertical direction.\\

Figure 2: The gain in magnetic energy $\varepsilon(\delta)-\varepsilon(0)$
as dimerization sets in with increasing $\delta$ for the four
configurations; (a) in the range $0 \leq \delta \leq 0.1$, and (b) in $0
\leq \delta \leq 1$.\\

Figure 3: Dependence of the energy gap parameter $D$ on $\delta $ for the
four dimerization configurations; (a) in the range $0\leq \delta \leq 0.1$,
and (b) in $0\leq \delta \leq 1$.

\newpage\ 

Table-1 : Summary of the critical exponents for spin-Peierls transition in a
Heisenberg chain determined by various methods.\\\flushleft{\ }

\begin{tabular}{||c|c|c|c||}
\hline\hline
Method & Interval & $\varepsilon (\delta )-\varepsilon (0)$ & Exponent \\ 
\hline
{Random phase app.$^{{\cite{cross}}}$} & $0\leq \delta \leq 1$ & $\delta ^x$
& $x=4/3$ \\ \hline
{Renormalization group$^{{\cite{fields}}}$} & $0\leq \delta \leq 1$ & $%
\delta ^x$ & $x=1.53$ \\ \hline
{2-level RG$^{{\cite{matsuyama}}}$} & $0\leq \delta \leq 1$ & $\delta ^x$ & $%
x=1.78$ \\ 
& $0.05\leq \delta \leq 0.1$ & {$\delta ^{2\nu }$}${/}${$\mid \ln $}${(}${$%
\delta $}${)}${$\mid $} & $2\nu =1.68_{-0.36}^{+0.13}$ \\ 
& $\ 0.4\leq \delta \leq 0.5$ & {$\delta ^{2\nu }$}${/}${$\mid \ln $}${(}${$%
\delta $}${)}${$\mid $} & $2\nu =1.31\pm 0.02$ \\ \hline
{Excitation spectrum$^{{\cite{bonner}}}$} & $0\leq \delta \leq 1$ & $\delta
^x$ & $x=1.36_{-0.2}^{+0.1}$ \\ \hline
Valence bond$^{\cite{soos1}}$ & $\delta \leq 0.05$ & {$\delta ^{2\nu }$}${/}$%
{$\mid \ln $}${(}${\ $\delta $}${)}${$\mid $} & $\nu =2/3$ \\ 
& $\delta \geq 0.05$ & $\delta ^x$ & $x=1.36_{-0.2}^{+0.1}$ \\ \hline
{Finit size scaling$^{{\cite{spronken}}}$} & $0\leq \delta \leq 0.1$ & {$%
\delta $}${^{2\nu }/}${$\mid \ln $}${(}${$\delta $}${)}${$\mid $} & $\nu
=0.71\pm 0.01$ \\ 
& $0\leq \delta \leq 1$ & $\delta ^x$ & $x=1.34\pm 0.02$ \\ \hline
{Exact diagonalization$^{{\cite{guo}}}$} & $0\leq \delta \leq 0.1$ & ${%
\delta ^{2\nu }/\mid \ln (\delta )}${$\mid $} & $\nu =2/3$ \\ \hline
DMRG$^{\cite{chitra}}$ & \multicolumn{1}{|c|}{$\delta \leq 0.05$} & $\delta
^x$ & $x=2/3$ \\ \hline\hline
\end{tabular}

\vspace{1.0cm}\ 

\end{document}